\documentstyle[11pt,aaspp4,flushrt]{article}
\newcommand\beq{\begin{equation}}
\newcommand\eeq{\end{equation}}

\begin{document}

\title{Anomalous X-ray Pulsars and Soft $\gamma$-ray Repeaters: Spectral
Fits and the Magnetar Model}

\author{Rosalba Perna,\altaffilmark{1} Jeremy S. Heyl,\altaffilmark{2}
  and Lars E. Hernquist}
\affil{\small Harvard-Smithsonian Center for Astrophysics, 60 Garden Street,
Cambridge, MA 02138; rperna@cfa.harvard.edu, jheyl@cfa.harvard.edu,
lhernqui@kona.harvard.edu}

\and

\author{Adrienne M. Juett and Deepto Chakrabarty}
\affil{\small Department of Physics and Center for Space Research,
  Massachusetts Institute of Technology, Cambridge, MA 02139;
  ajuett@space.mit.edu, deepto@space.mit.edu}

\altaffiltext{1}{Harvard Junior Fellow}
\altaffiltext{2}{{\em Chandra} Fellow}

\begin{abstract}
The energy source powering the X-ray emission from anomalous X-ray
pulsars (AXPs) and soft $\gamma$-ray repeaters (SGRs) is still
uncertain.  In one scenario, the presence of an ultramagnetized
neutron star, or ``magnetar'', with $B \simeq 10^{14} - 10^{15}$ G is
invoked.  To investigate this hypothesis, we have analyzed archival
{\em ASCA} data for several known AXPs and SGRs, and fitted them with
a model in which all or part of the X-ray flux originates as thermal
emission from a magnetar.  Our magnetar spectral model includes the
effects of the anisotropy of the heat flow through an ultramagnetized
neutron star envelope, reprocessing by a light element atmosphere, and
general relativistic corrections to the observed spectrum.  We obtain
good fits to the data with radii for the emitting areas which are generally
consistent with those expected for neutron stars, in contrast to 
blackbody (BB) fits, which imply much smaller radii. Furthermore, the
inclusion of atmospheric effects results in inferred temperatures which
are lower than those implied by BB fits, but however still too high to
be accounted by thermal cooling alone. An extra source of heating
(possibly due to magnetic field decay) is needed.  Despite the harder
tail in the spectrum produced by reprocessing of the outgoing flux
through the atmosphere, spectral fits still require a considerable
fraction of the flux to be in a power-law component. 
\end{abstract}

\keywords{stars: neutron --- $X$-rays: stars}

\section{Introduction}

The roughly half dozen so-called anomalous X-ray pulsars (AXPs;
Mereghetti \& Stella 1995; van Paradijs, Taam, \& van den Heuvel 1995)
have recently emerged as a distinct class of objects.   They are slow
($P\sim 5$--10~s) rotators with no evidence of Doppler shifts from
binary motion, and several are associated with supernova remnants,
suggesting that they are young objects.   Owing to their long periods,
their rotational energy loss is far too 
low to power their observed X-ray luminosities.  Models to account for the
properties of these sources fall into two broad categories. In the
first class of models, the X-ray emission is powered by accretion,
which could result from a binary companion of very low mass (Mereghetti \&
Stella 1995), the debris of a disrupted massive companion (van
Paradijs, Taam \& van den Heuvel 1995; Ghosh, Angelini \& White 1997),
or material falling back after a supernova explosion (Corbet et
al. 1995; Chatterjee, Hernquist \& Narayan 2000; Alpar 1999, 2000;
Marsden et al. 2001).  In the second class of models, accretion is not
involved; instead, the AXPs are hypothesized to be either 
ultramagnetized neutron stars (NSs; Thompson \& Duncan 1996) or remnants of
Thorne-${\dot {\rm Z}}$ytkow objects (van Paradijs et al. 1995). In
the former case, the X-ray luminosity could be powered either by
magnetic field decay (Thompson \& Duncan 1996; Heyl \& Kulkarni 1998)
or by residual thermal energy (Heyl \& Hernquist 1997a,b).

If the AXPs are indeed ultramagnetized NSs (or ``magnetars''; Duncan
\& Thompson 1992, 1995), then they may be related to another class of
objects, the soft $\gamma$-ray repeaters (SGRs; e.g. Kouveliotou et 
al. 1999).  In their quiescent X-ray emission, the SGRs have slow
pulsations similar to those seen in the AXPs.  In addition, the SGRs
sometimes show strong hard X-ray/soft gamma-ray bursts, which can be
distinguished from the classical $\gamma$-ray bursts (GRBs) by their recurrence
and their spectra, which are generally much softer than those of the GRBs.  
According to Thompson \& Duncan
(1996), these bursts could be the result of sudden releases of energy
resulting from rearrangement of the magnetic fields
in the crusts of highly magnetized NSs.

In the magnetar model for the AXPs and SGRs, whether the X-ray
luminosity is powered by cooling or magnetic field decay or by
a combination of both, the thermally emitting area must be consistent with 
a neutron star surface. However, blackbody fits to the spectra of
these objects require emitting areas that are only a small fraction of
the total surface. This discrepancy may arise at least partially
because the thermal emission of NSs most likely does not have a true
blackbody spectrum.  Atmospheric effects are known to distort NS
spectra.  Fallback after the supernova explosion, and/or
accretion from the interstellar medium, will likely cover the surface of the
star with light elements. (Note that to cover the surface to an X-ray
optical depth of unity requires only $\sim 10^{14} {\rm g} -
10^{-19} M_\odot$ of material.) Moreover, this material is likely to
suffer significant fractionization on short timescales (Alcock \& Illarionov 1980;
Romani 1987).   Owing to the
enormous surface gravity of the star, the heavier elements will
settle out in of order 1$-$100 sec, leaving an atmosphere made of
light elements. Besides atmospheric effects, the emergent NS spectrum
is also dependent on the underlying temperature distribution on the
stellar surface (in blackbody fits this is  assumed incorrectly to be 
uniform), 
and
on the general relativistic effect of light deflection owing to the
large surface gravity. A spectral analysis that takes into account all
these effects is the purpose of this work.

More specifically, in this {\em Letter}, we analyze archival {\em
ASCA} data for some of the known AXPs and SGRs, and fit them with a
model in which the X-ray flux is produced by thermal emission from a
highly magnetized NS with an atmosphere made of light elements. We
take into account the anisotropy in the flow of heat through the
envelope of the NS owing to the intense magnetic field, as well as
general relativistic corrections to the observed spectrum.  Our fits
are consistent with emission from the entire surface of a neutron
star, supporting the interpretation of AXPs and SGRs as magnetars.
  
\section{The X-Ray Spectrum of a Cooling Magnetar}

We consider a highly magnetized neutron star cooling through an
accreted envelope. Heyl \& Hernquist (1998b, 2000) showed that, if
$B_p\ga 10^{12}$ G, the flux transmitted through the envelope can be
well approximated by $F\propto \cos^2\psi$, where $\psi$ is the angle
between the local radial direction and the magnetic field.  For a dipolar field, 
\beq
\cos^2\psi=4\cos^2\theta_p/(3\cos^2\theta_p+1)
\label{eq:cospsi}
\eeq
(Greenstein \& Hartke 1983), where $\theta_p$ is the angle between the
radial direction at position ($\theta,\phi$) on the
surface of the star, and the magnetic pole.  In spherical coordinates, it
is given by
\beq
\cos\theta_p=\cos\theta\cos\alpha+\sin\theta\sin\alpha\cos\phi\;,
\label{eq:tetap}
\eeq where $\alpha$ is the angle that the magnetic pole makes with the
line of sight.  For this study, we consider an orthogonal
rotator.\footnote{We find that spectral fits are insensitive to this
choice when considering the average flux over the rotational period of the
star.}  The angle $\alpha$ is then coincident with the phase angle $\Omega t$,
where $\Omega$ is the angular velocity of the star.  For the local
emission from the NS surface, $n(E,T)$, we assume a blackbody spectrum
modified by the presence of an atmosphere made of light elements, for
which we adopt the semianalytical model of Heyl \&
Hernquist\footnote{This model yields spectral intensities very close
to those of Pavlov et al. (1994), who computed detailed atmospheres
for magnetic field strengths on the order of $10^{12}-10^{13}$~G.
Preliminary results (D. Lloyd et al.  in preparation) suggest
furthermore that realistic spectra are similar to ours even for much
stronger fields, $B\sim 10^{14}-10^{15}$ G. } (1998a), but with the
inclusion of limb darkening.  This is parameterized by an angle
dependence of the intensity $\propto \cos^\beta\delta$. The dependence
on $\beta$ is explored in the fits.

Let $R$ be the radius of the NS star, $M$ its mass (for which we adopt
$M=1.4 M_\odot$), $R_s=2GM/c^2$ its Schwarzschild radius, and define
$e^{-\Lambda_s}\equiv\sqrt{1-R/R_s}$.  If $D$ is the distance from the
star to the observer, the phase-averaged flux measured by an observer
at infinity (without including the reduction due to photoelectric
absorption by intervening material) is given by
\beq
f(E)=\frac{\pi R_\infty^2\;\sigma T^4_{p,\infty}} 
{4\pi D^2}\frac{1}{k T_{p,\infty}}
\int_0^{2\pi}\frac{d\alpha}{2\pi}\int_0^1 2xdx\int_0^{2\pi}
\frac{d\phi}{2\pi}\; 
I_0(\theta,\phi) \;n[Ee^{-\Lambda_s};T_{s}(\theta,\phi)]\;,
\label{eq:flux}
\eeq 
in units of photons cm$^{-2}$ s$^{-1}$ keV$^{-1}$.  Here $x=\sin\delta$
($\delta$ being the angle between the normal to the NS surface and the
direction of the photon trajectory), $R_\infty\equiv Re^{\Lambda_s}$,
and $T_{p,\infty}\equiv T_{p}e^{-\Lambda_s}$, where $T_p$ is the
temperature at the pole.  The general relativistic effects of light
deflection are taken into account through the ray-tracing function
(Page 1995)
\beq
\theta(\delta)=\int_0^{R_s/2R}x\;du\left/\sqrt{\left(1-\frac{R_s}{R}\right)
\left(\frac{R_s}{2R}\right)^2-(1-2u)u^2 x^2}\right.\;,
\label{eq:teta}
\eeq
where $u=R_s/2r$, with $r$ being the radial coordinate.
A photon emitted at an angle $\delta$
with respect to the normal to the surface comes from a colatitude
$\theta(\delta)$ on the star.  The total flux at each point
$(\theta,\phi)$ of the NS surface is given by
\beq
I_0(\theta,\phi)=\frac{4\cos^2\theta_p}{3\cos^2\theta_p+1}\;
(0.75\;\cos^2\theta_p+0.25)^{0.2}\;.
\label{eq:I0}
\eeq 
The first term in the right hand side is appropriate for a dipole, and
we have then assumed a further dependence of the flux on $B^{0.4}$, as
in Heyl \& Hernquist (1998b).  Finally, the local temperature on the 
stellar surface is
determined by 
\beq
T_{s}(\theta,\phi)=T_p[I_0(\theta,\phi)]^{1/4}\;. 
\label{tstar}
\eeq

\section{Observations and Analysis}

We analyzed archival observations of several AXPs and SGRs made by the {\em
Advanced Satellite for Cosmology and Astrophysics} ({\em ASCA};
Tanaka, Inoue, \& Holt 1994), obtained from the High Energy
Astrophysics Archival Research Center (HEASARC) at NASA Goddard Space
Flight Center.  {\em ASCA} was launched in 1993 and continued to make
observations through 2000.  It carried four identical grazing
incidence X-ray telescopes capable of imaging X-rays in the 0.5--10
keV range with a 24 arcmin (FWHM) field of view, a $\sim$1 arcmin
point spread function, and a total effective area of 1300~cm$^2$ at 1
keV.  Each telescope had one dedicated focal plane instrument, and all
four instruments simultaneously recorded data for each observation.  
The two CCD cameras (SIS0 and SIS1) had superior energy resolution
($E/\Delta E\sim$20--50) and were sensitive down to about 0.5~keV.
The two gas scintillation imaging proportional counters (GIS2 and
GIS3) had more modest energy resolution ($E/\Delta E\sim$10) and 
little sensitivity below 1 keV, but their effective area was
comparable to the CCDs around 2 keV and higher above $\sim$4 keV.  
Thus, the GIS data were somewhat better suited for studying the
X-ray continuum spectrum of absorbed sources like the AXPs and SGRs. 
However, we initially examined data from all four instruments in our
analysis.  

A summary of the {\em ASCA} observations that we analyzed is given in
Table~1.  For simplicity, we confined ourselves to those sources where
there is no contaminating emission from a surrounding supernova
remnant to consider; this excludes the AXPs 1E 2259+586 and 1E 1841-045.  We used the 
standard screened events files provided through HEASARC by the {\em ASCA} 
Guest Observer Facility (GOF).  From these files, spectra were extracted
for the point sources.  The extraction radii were 6 arcmin and 4arcmin for GIS and 
SIS respectively.  Background spectra were also extracted from the 
event files from the area outside a 8 arcmin radius from the point source and 
any other bright sources present in the observation.  We fitted the 
X-ray spectra for all the observations in Table~1 using the XSPEC 
spectral analysis package (Arnaud 1996).  In each case, we fit the 
analytic magnetar model described in Equation~(\ref{eq:flux}).  For 
comparison, we also fit an ideal blackbody model (BB), assuming a uniform 
temperature distribution over the emitting area.  The free parameters of 
the magnetar model are the pole temperature $T_p$ and the radius of the
star $R$. (For a non-relativistic star, the radius affects only the
overall normalization of the spectrum; however, when general
relativistic effects are taken into account, the radius also modifies
the shape of the spectrum through the dependence of the function
$\theta(x)$ in Equation~(\ref{eq:flux}) on $R$.)  

For both the magnetar and the BB models, we included the
multiplicative effect of interstellar photoelectric absorption
(Morrison \& McCammon 1983).  We also allowed for the possibility of
an absorbed power law (PL) component, using the same absorption column
as for the thermal component.  However, for each object, a fit using
only the absorbed thermal component alone was also attempted.  A
spectrum produced by an atmosphere indeed has a harder tail compared
to simple blackbody emission at the same effective
temperature. Therefore, even if the blackbody fit always required an
extra power law component, this did not necessarily have to be the
case when processing by an atmosphere was included in the computation
of the spectrum.

For each observation, we began by fitting the data for each instrument
individually, in order to evaluate the data quality separately.  If the
data from one or more of the instruments had poor statistics (usually
in the case of the SIS data) or was otherwise problematic, it was
discarded.  The remaining data were then fit jointly to maximize our
continuum sensitivity, with the overall normalization tied to that of
the GIS2 detector (which generally gave the best independent fits).  
In the case of the SGR 1627$-$41 and SGR 1806$-20$, we found that
there were too few soft counts to allow a meaningful constraint on the
fit parameters for the thermal components.  We do not consider these
sources further in this paper.  

The results of our fits are summarized in Table 2.  For each source,
we show both the absorbed BB+PL fits and the absorbed magnetar+PL
fits.  In the case of AX J1845-0258, however, the statistics were too poor
to constrain both components separately. For this object, the thermal
and non-thermal components in Table 2 represent alternate fits to
the same data.  In a few cases (i.e. when the fit was acceptable), we
also show an absorbed magnetar-PL fit with the column density held
fixed at $N_{\rm H}=1\times 10^{22}$ cm$^{-2}$.  For all our fits, 
we also give
the fraction of the flux that is in the power law
component, computed in terms of the unabsorbed photon flux in the
0.7--10 keV band.  A comparison between the temperatures and radii
obtained in the BB fits and in the fits with the atmospheres is shown
in Figure 1 (except for the object RXS J1708$-$40, whose distance is still
largely uncertain, and for which a separate discussion will be made in
\S 4).  The two dashed lines mark the region of NS radii which are
allowed by currently available models for the NS equation of state.

\section{Discussion}

We have analyzed archival {\em ASCA} data for several known AXPs and SGRs and
fitted their spectra with a model in which the X-ray emission
consists of thermal radiation from a highly magnetized neutron star,
as well as a power-law component at high energies.  For the thermal
contribution, we have included distortions in the spectra due to the
presence of an atmosphere of light elements, accounted for the
anisotropic flow of heat through the envelope due to the intense
magnetic field, and included general relativistic corrections to the
observed flux.  We find that the thermal emitting areas implied by our
model are always larger than those derived by spectral fits that use a
blackbody spectrum, and are generally consistent with those expected
for neutron stars.  All the fits were made using a model with a
moderate beaming, i.e.  $\beta\sim 0-1$.  We have considered other
radiation patterns and find that a more intense beaming results in
slightly larger inferred areas and smaller temperatures, but the fits
are equally acceptable.  Therefore, spectral fits alone made with the
phase averaged spectrum are not able to constrain the degree and type
of beaming, especially given the uncertainties in the other
parameters.

In most of the cases we found that, despite the fact that processing
by an atmosphere leads to lower temperatures than those implied by the
BB fits, our inferred temperatures are still too high to be accounted
for by thermal cooling alone.  However, if the magnetic field is
sufficiently strong and decaying, the energy from its decay may
augment the thermal emission from the surface (Heyl \& Kulkarni 1998).
The inferred surface temperatures of all of the objects with well
constrained fits are higher than would be expected for a neutron star
cooling through an iron envelope (Heyl \& Hernquist 2000); the results
for the AXP 4U 0142$+61$, RXS J1708$-$40 and SGR 1900$+$14
are marginally consistent with a neutron star cooling through a highly
magnetized iron envelope with a substantial contribution to the flux
from the decay of the magnetic field (Heyl \& Kulkarni 1998).  If the
energy released by the magnetic field is deposited at high densities,
magnetic field decay alone cannot explain the high effective
temperatures of the AXPs 1E 1048.1$-5937$ and 1E 1845$-$0245.  Thermal
emission through a light element envelope (Heyl \& Hernquist 1997b,
Potekhin, Chabrier \& Yakovlev 1997), however, can account for these
sources. 

Atmospheric effects play a crucial role in determining the emitted
spectrum (Romani 1987; Pavlov et al. 1994; Zampieri et al. 1995;
Rajagopal \& Romani 1996; Zavlin, Pavlov \& Shibanov 1996).  For the
effective temperatures of these stars (i.e. $\sim$ a few $\times 10^6$
K), the spectra from hydrogen atmospheres depend only weakly on the
strength of the magnetic field (see e.g.  Fig. 5 in Rajagopal, Romani
\& Miller 1997).  However, the composition of the atmosphere may
dramatically affect the emitted spectrum (see, e.g. references above).
If the atmosphere were made of heavy elements such as iron, the
emitted spectrum would be much closer to a blackbody\footnote{Note
that the accretion model predicts that the opacity should be dominated
by heavy metals in SGRs and AXPs, because the settling time is
dominated by the rate of metal deposition for the accretion rates
implyed by the X-ray emission (Brown, Bildsten \& Rutledge 1998).}
(Rajagopal et al. 1997).

Note that, given the mild dependence of the spectrum on the strength
of the magnetic field for $B\ga 10^{12}$ G, the model we have
developed could similarly be used to model the thermal emission
of neutron stars with magnetic fields $\sim 10^{12}- 10^{13}$ G and
light element atmospheres. However, in the case of the AXPs, the
additional heat (with respect to the predictions of standard cooling
scenarios) would be difficult to explain without the contribution from
magnetic field decay. A hypothetical contribution from accretion
would most likely result from matter channeled onto the magnetic
poles of the star by the presence of
the magnetic field lines. This would yield hot polar caps and the
total spectrum would show two thermal components, one at lower energy
being consistent with the total area of the star (due to thermal
cooling) and another at higher temperature but coming from a small
fraction of the star (due to the heated polar caps). Here
we find that spectral fitting requires only one thermal
component, and therefore this scenario does not seem to be
favored. However, it is also true that the high column density to
these sources would make a low-temperature thermal component difficult
to detect.

The distances to the sources are still rather uncertain. As Figure 1
shows, relatively small variations with respect to the assumed values
would not affect any of the conclusions of this work. The value of the
inferred radius, $R$, roughly scales with the distance $D$. If any of
the distances turned out to be much smaller than the assumed value, so
that the radii were consequently much smaller than the
minimum allowed value by the NS equation of state, then an accretion
model where the X-ray emission is produced by a hot spot would be
favored with respect to the magnetar model, where it is produced by
the whole surface of the star.  On the other hand, if any of the
distances turned out to be much larger than the assumed value, so that
the corresponding BB radius were already consistent with the emission
from the entire surface, then a light element atmosphere would not
be appropriate, as it would yield too large radii, while
an atmosphere made of heavy elements such as iron would be a viable
option, because it yields spectra very close to blackbody, as
discussed above. In such a case, it would be difficult to interpret
the X-ray emission as being powered by accretion unless the magnetic
field of the objects were very small, so that accretion could proceed
in a quasi-spherical fashion rather than being channeled through the
poles\footnote{Note that in models where accretion derives from
fallback disks (e.g. Chatterjee et al. 2000), a magnetic field of the
order of $10^{12}-10^{13}$ G is needed.}.

Among all the objects that we considered, RXS J1708$-40$ is the one
which has the largest uncertainty in its distance. It lies in the
Galactic plane, and, along this direction, spiral arms are located at
a distance of $\sim 1$, $\sim 3$, and $\sim 4.5$ kpc (Taylor \& Cordes
1993).  The large column density to the source inferred from the X-ray
spectra suggests a likely distance in the 5-10 kpc range (Israel et
al. 1999).  However, a smaller distance cannot be ruled out. Our
results suggest that this object, if it were a magnetar with a light
element atmosphere, would most likely reside in the middle spiral
arm. However, a magnetar with an atmosphere of heavy elements would be
consistent with the largest estimates of the distance.

Besides the distance to the sources, the main uncertainty in the
inferred emitting areas from the spectral fits arises from the
uncertainty in the column density $N_{\rm H}$. This 
conclusion is similar to 
the findings of Rutledge et al. (1999) for their spectral fits of the
quiescent X-ray emission from accreting NS transients. This uncertainty may
be reduced if high signal-to-noise spectral data becomes available in a
passband that covers both the energy range where absorption is most
effective and where it is not.  Such observations may be possible
with {\em Chandra X-Ray Observatory} and {\em XMM-Newton}.

Despite the harder tail in the spectrum produced by reprocessing of
the outgoing flux through an atmosphere, spectral fits still require
in most cases a hard power law component
 (even though its normalization is
often not well constrained).  In the context of the magnetar model,
the origin of this power-law emission is not fully understood.
Thompson \& Duncan (1996) have proposed that the non-thermal
components of AXP and SGR spectra can be produced by magnetospheric
currents resulting from fracturing of the neutron star crust.  An
important constraint on this type of model is provided by the fact
that the hard emission from AXPs appears to be pulsed, often with a
large amplitude.  Although the angular distribution of the radiation
from these effects is not yet known, a magnetospheric origin for the
power-law component is appealing from the point of view of helping to
account for the observed pulsations, because gravitational bending
would be relatively less significant far from the stellar surface.  A
preliminary analysis (C. Thompson, private communication) suggests
that photons produced by magnetospheric currents will preferentially
escape from regions near the magnetic poles, possibly resulting in a
large amplitude of pulsation. However, in order to be able to draw
firmer conclusions from a timing analysis, tighter limits on the power
law fraction $f_{\rm pl}$ are needed, possibly as a function of
energy. To such purpose, independent constraints on the column
densities would be highly desirable.  In the fits in which $N_{\rm
H}$ is held fixed the normalization of the power law component is much
better constrained. Moreover, the column density itself influences the
inferred values of the pulsed fractions (Perna, Heyl \& Hernquist
2000).

In summary, our finding that plausible atmosphere models yield thermal
emitting areas consistent with a neutron star surface supports the
interpretation of AXPs and SGRs as magnetars.  However, based on
currently available data, we cannot definitively discriminate between
the magnetar and accretion models for these objects.  In particular,
the apparent requirement that a large fraction of the flux arises from
a hard power-law component significantly complicates efforts to infer
the true nature of these sources.

A variety of objections have been raised against both the magnetar and
accretion models for AXPs and SGRs (e.g. Li 1999; Marsden, Rothschild
\& Lingenfelter 1999; DeDeo, Psaltis \& Narayan
2000; Hulleman et al. 2000a; Marsden et al. 2001). In the case of the magnetar
interpretation, our results show that the thermal emission is
consistent with the surface area of a neutron star, provided that the
spectrum from this component is sufficiently distinct from a
blackbody.  Moreover, the existence of a substantial non-thermal,
power-law component mitigates concerns that the large pulsed fractions
measured for some AXPs and SGRs might be inconsistent with the
magnetar hypothesis.  For an object like 1E 1048.1-5937, for example, the
unusually large pulsed fraction of $\sim 70\%$ (e.g. Corbet \& Mihara
1997; Oosterbroeck et al. 1998) may simply reflect the inferred
presence of a particularly important non-thermal process whose
contribution in this case approaches $\sim 80\%$ of the total flux.
It remains problematic, however, whether or not models
for the origin of the power-law emission in the context of
the magnetar model can account for such high relative fluxes.

Perhaps the most challenging argument against the accretion model for
AXPs and SGRs comes from recent studies of optical and infrared
emission from these objects.  In the work of Chatterjee et al. (2000),
for example, the presence of an extended disk is expected to yield
significant optical and infrared fluxes (e.g. Perna et al. 2000; Perna
\& Hernquist 2000), in disagreement with existing observational
measurements (e.g. Hulleman et al. 2000a,b).  However, Menou et
al. (2001) have recently proposed an alternate accretion scenario
based on fallback in which the disk always remains geometrically thin
and radially compact and accretion is halted on timescales comparable
to the ages of AXPs and SGRs through a thermal ionization instability.
In this model, the optical and infrared emission from the disk is
greatly suppressed relative to that predicted by e.g. Chatterjee et
al.  since the disk is limited in radial extent.  
    
Another argument against the accretion model for AXPs concerns the
unusually steady spindown of objects like RXS J1708$-40$ and 1E 2259$+586$
(Kaspi, Chakrabarty \& Steinberger 1999).  Such behavior is not
characteristic of accreting, binary X-ray pulsars, but the
implication of this finding for isolated fallback disks is less
clear, particularly for models in which the disks are low in mass
(Menou et al. 2001) and considering the noisier behavior of AXP
1E 1048.1$-$5937 (Kaspi et al. 2000).

Thus, it appears that with existing observational constraints it is
not possible to conclusively rule out either the magnetar or accretion
scenario for AXPs and SGRs.  In detail, however, the spectral energy
distributions expected in these two classes of models should exhibit
significant differences in many wavelength intervals.  Hence, we are
optimistic that highly sensitive ongoing multi-wavelength observations
combining X-ray spectroscopy with deep optical and infrared searches
will eventually discriminate between the magnetar and accretion
pictures, definitively revealing the true nature of these sources.

\acknowledgements{We thank Chris Thompson for enlightening discussion,
Jonathan McDowell for support with the XSPEC software, and an anonymous
referee for insightful comments.
Support for JSH was provided by the National Aeronautics and
Space Administration through Chandra Postdoctoral Fellowship Award
Number PF0-10015 issued by the Chandra X-ray Observatory Center, which
is operated by the Smithsonian Astrophysical Observatory for and on
behalf of NASA under contract NAS8-39073.
}

\clearpage

\begin{deluxetable}{llc}
\tablewidth{0pt}
\tablecaption{{Log of \em ASCA} Observations Analyzed}
\tablehead{\colhead{Source} & \colhead{Date} & \colhead{Exposure (ks)}}
\startdata
\cutinhead{Anomalous X-ray pulsars}
4U 0142$+$61 & 1998 August 21 & 37.7 \\
1E 1048.1$-$5937 & 1998 July 26 & 117.7 \\
RXS J1708$-$40 & 1996 September 3 & 33.7 \\
AX J1845$-$0258  & 1993 October 12 & 89.0 \\
\cutinhead{Soft $\gamma$-ray repeaters}
SGR 1627$-$41 & 1999 February 26 & 261.6 \\
SGR 1806$-$20 & 1993 October 10 & 108.7 \\
              & 1993 October 20 & 133.3 \\
SGR 1900$+$14 & 1998 September 16 & 174.8 \\
\enddata
\end{deluxetable}

\clearpage
\begin{deluxetable}{lccccccc}
\tablewidth{0pt}
\tablecaption{Spectral Fits}
\tablehead{
  & & \multicolumn{2}{c}{Thermal component} & 
  \multicolumn{2}{c}{Power law component} &  & 
\\ \cline{3-4} \cline{5-6}
  \colhead{Model} & \colhead{$N_{\rm H}$ ($10^{22}$ cm$^{-2}$)} & 
  \colhead{$kT$ (keV)} & \colhead{$R$ (km)} & \colhead{$\gamma$} & 
  \colhead{$C_1$\tablenotemark{a}} & 
\colhead{$\chi^2_{\rm red}$/$N_{\rm dof}$}&\colhead{$f_{\rm pl}$\tablenotemark{b} $\;$(\%)} }
\startdata
\cutinhead{{\em 1E 1048.1$-$5937 (AXP)}, $d=10$ kpc\tablenotemark{(c)}}
BB+PL  & 1.47$_{-0.2}^{+0.14}$ &  0.63$_{-0.04}^{+0.03}$ &
      1.8$_{-0.2}^{+0.3}$ & 3.9$_{-0.3}^{+0.5}$ & 
      14$_{-10}^{+20}$&       0.89/771 & 80\\
Magnetar+PL   &  1.62$_{-0.25}^{+0.34}$&  0.43$_{-0.02}^{+0.04}$&  15.2$_{-0.7}^{+4}$&
4.9$_{-0.8}^{+1.2}$&     20$_{-9}^{+23}$&     0.89/771 & 78\\
Magnetar+PL   &  1.0 (fixed)&  0.41$_{-0.01}^{+0.01}$&   16.5$_{-0.9}^{+1.3}$&
3.4$_{-0.3}^{+0.3}$&      3.7$_{-0.5}^{+0.4}$&   0.91/772 & 45\\
\cutinhead{{\em 1E 1845$-$0258 (AXP)}, $d=8.5$ kpc\tablenotemark{(d)}}
BB  & 5.16$_{-0.96}^{+1.08}$&  0.66$_{-0.06}^{+0.06}$&  2.0$_{-0.5}^{+0.8}$&
    &      &       0.94/177 & 0\\
PL only   &  10.3$_{-1.5}^{+1.8}$&  &  &
5.1$_{-0.6}^{+0.7}$&     974$_{-828}^{+43}$&     1.00/177 & 100\\
Magnetar   &  6.3$_{-1.3}^{+2.1}$ &  0.41$_{-0.09}^{+0.07}$&   18$_{-7.2}^{+26}$&
&     &   0.97/177 & 0\\
\cutinhead{{\em 4U 0142$+$61 (AXP)}, $d=1$ kpc\tablenotemark{(e)}}
BB+PL  & 1.12$_{-0.08}^{+0.08}$&  0.41$_{-0.01}^{+0.02}$&  1.7$_{-0.3}^{+0.2}$&
4.0$_{-0.1}^{+0.2}$&      333$_{-60}^{+64}$&       0.97/330 & 88\\
Magnetar+PL   &  0.62$_{-0.14}^{+0.3}$&  0.28$_{-0.01}^{+0.01}$&
16.1$_{-0.6}^{+0.9}$&  3.2$_{-0.6}^{+0.7}$&     44$_{-31}^{+130}$& 0.98/330 &35\\
Magnetar+PL   &  1.0 (fixed)&  0.27$_{-0.01}^{+0.01}$& 14.4$_{-1.1}^{+1.0}$&
4.0$_{-0.08}^{+0.06}$&      216$_{-10}^{+15}$&   0.99/331 & 75\\
\cutinhead{{\em RXS J1708$-$40 (AXP)}, $d=10$ kpc\tablenotemark{(f)}}
BB+PL  & 1.47$_{-0.46}^{+0.5}$&  0.40$_{-0.08}^{+0.06}$&  11.5$_{-7}^{+6}$&
2.8$_{-1}^{+0.7}$&      36.2$_{-30}^{+80}$&       0.82/128 & 73\\
Magnetar+PL   &  1.24$_{-0.33}^{+0.66}$&  0.31$_{-0.11}^{+0.05}$&
69$_{-33}^{+75}$&  2.1$_{-2}^{+1.4}$&     8.2$_{-8}^{+95}$& 0.93/128 &35\\
Magnetar+PL   &  1.0 (fixed)&  0.34$_{-0.01}^{+0.02}$& 56$_{-8}^{+4}$&
1.3$_{-1.5}^{+0.8}$&      1.7$_{-1.6}^{+7}$&   0.82/129 & 25\\
\cutinhead{{\em SGR 1900$+$14}, $d=5$ kpc\tablenotemark{(g)}}
BB+PL  & 2.5$_{-0.3}^{+0.2}$&  0.53$_{-0.16}^{+0.08}$&  1.2$_{-0.7}^{+0.8}$&
2.1$_{-0.2}^{+0.3}$&      105$_{-20}^{+80}$&       0.83/309 & 90\\
Magnetar+PL   &  2.4$_{-0.3}^{+0.4}$&  0.34$_{-0.16}^{+0.05}$&  11.3$_{-0.3}^{+1.4}$&
2.1$_{-0.3}^{+0.3}$&     110$_{-100}^{+70}$&     0.83/309 & 85\\
\enddata
\tablenotetext{a}{Power-law normalization at 1 keV in units of
$10^{-3}$ photons cm$^{2}$ s$^{-1}$ keV$^{-1}$.}
\tablenotetext{b}{Energy flux in the power law component for the 0.7-10 keV band.}
\tablenotetext{(c)}{Van Paradijs et al. (1995).}
\tablenotetext{(d)}{Torii et al. (1998).}
\tablenotetext{(e)}{Israel et al. (1999), but see also text.}
\tablenotetext{(f)}{White et al. (1996).}
\tablenotetext{(g)}{Hurley et al. (1999).}

\end{deluxetable}

\clearpage

\begin{figure}[t]
\centerline{\epsfysize=5.7in\epsffile{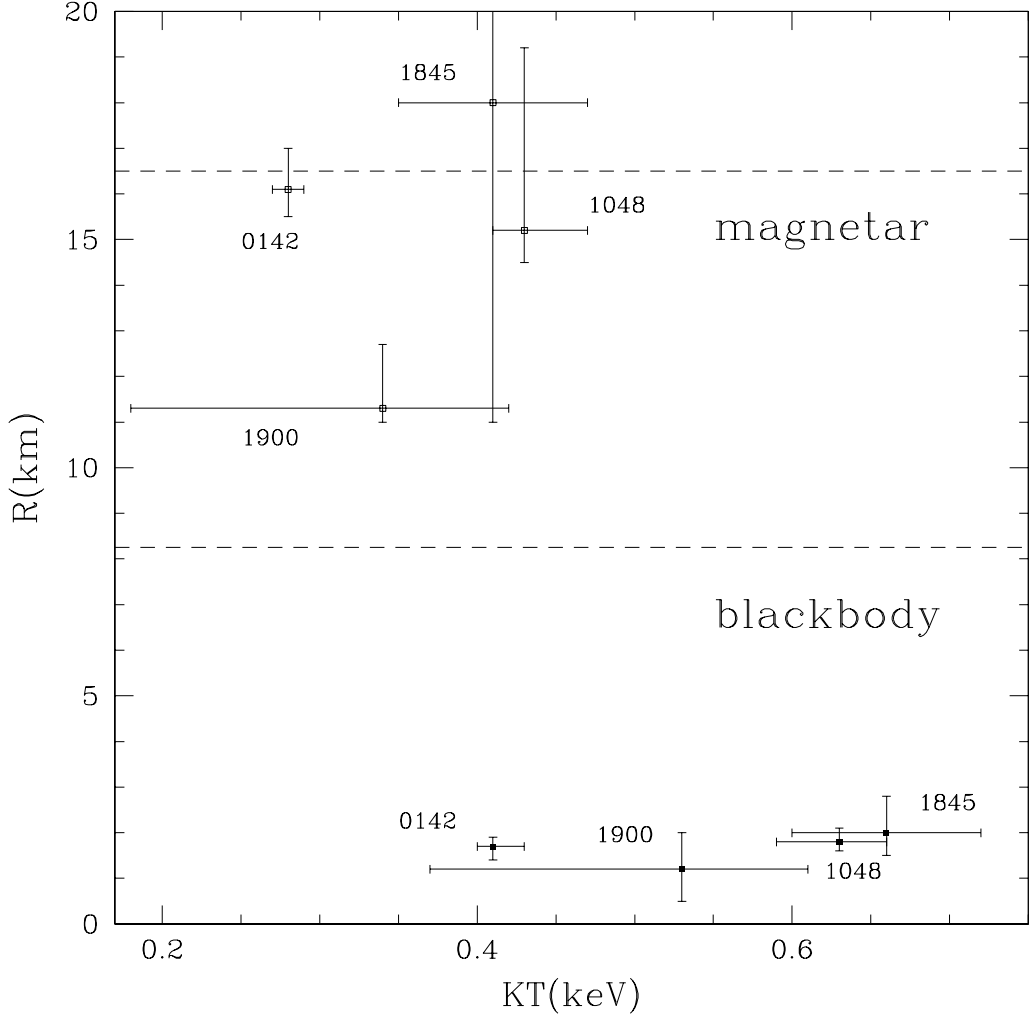}}
\caption{Inferred temperatures and radii for blackbody fits, and for fits
done with the magnetar model. The two dashed lines mark
the region of NS radii which are allowed by currently available models
for the NS equation of state for a NS of $M=1.4M_\odot$
(Pandharipande 1971; Pandharipande \& Smith 1975) .}
\label{fig:1} 
\end{figure}

\end{document}